\documentclass[12pt,a4paper]{article}
\usepackage{array,enumerate,isolatin1}
\usepackage{epsfig,rotating,pstricks,colordvi,fancyheadings}
\makeatletter
\renewcommand{\@seccntformat}[1]{{\csname the#1\endcsname}\mbox{.  }}
\makeatother
\begin{document}
\renewcommand \arraystretch{1.3}
\setcounter{secnumdepth}{1}
\thispagestyle{empty}
\pspicture(0,-1)(0.1,0.1)
\rput{0}(10.0,1){\footnotesize  submitted to Surf.~Rev.~Lett.}
\endpspicture
\vspace{-1cm}
\begin{center}
{\bf \Large Theory of Alkali Induced Reconstruction of the Cu(100) Surface}\\[3mm]
{\Large S.~Quassowski and K.~Hermann} \\[3mm]
{Fritz--Haber--Institut der Max-Planck-Gesellschaft,
Faradayweg 4--6, D--14195 Berlin, Germany} \\[3mm]
\end{center}
\begin{abstract}
LEED experiments show that Li adsorbed at Cu(100) surfaces at room
temperature induces a (2$\times$1) missing row substrate reconstruction 
while adsorption at lower
temperatures, \mbox{$T$=180 K}, results in an unreconstructed Cu(100)+c(2$\times$2)--Li
overlayer structure. Substrate reconstruction has not been observed for Na
nor for K adsorption. In order to study the specific 
reconstruction behavior of the Li
adsorbate ab initio DFT calculations have been performed on \mbox{Cu(100)+Ad},
\mbox{Ad = Li, Na, K} systems at coverages $\Theta_{\mathrm{Ad}}=0.25 - 0.5$ with and
without reconstruction. The calculations show that the (2$\times$1) MR
reconstructed surface lies energetically above the ideal (1$\times$1) surface by
\mbox{$0.2$ eV} per unit cell. However, alkali binding is stronger in the MR
geometry as compared to that of the ideal surface where the increase in
bond strength becomes smaller in going from Li to Na to K. As a result,
the MR reconstructed and the overlayer adsorbate systems are energetically
very close for \mbox{Cu(100)+Li} while for Na and K the overlayer geometry is
always favored.
\end{abstract}

\section{Introduction}
It is well known from experiment that alkali adsorption at fcc(110)
surfaces induces a (2$\times$1) missing row (MR) substrate reconstruction (for
reviews see \cite{Ove95,Die95}). Recently, this type of adsorbate induced
reconstruction was also found on fcc and bcc (100) surfaces. Examples are
Au(100)+K \cite{Oka92}, Au(100)+Na \cite{Neu95}, Ag(100)+K \cite{Oka91},
and Cu(100)+Li \cite{Toc92,Miz93}. Here the Cu(100)+Li system is of
particular interest since it exhibits different reconstruction behavior
depending on the adsorption temperature \cite{Toc92,Miz93}. At low
temperatures, $T \le 180 K$, a c(2$\times$2) overlayer (OL) structure is observed for
$\Theta_{\mathrm{Li}} \approx 0.5$ without indications of a substrate
reconstruction. However,
subsequent heating of the adsorbate system to room temperature leads to a
(2$\times$1)MR reconstruction of the Cu(100) substrate which is conserved even
if the system is cooled back to $180$ K. This type of adsorbate induced
substrate reconstruction seems unique for Li adsorption on Cu(100) and has
not been found for larger alkali adsorbates such as Cu(100)+Na
\cite{Rei91}, Cu(100)+K \cite{Aug85,Mey94}, or Cu(100)+Cs \cite{Mey94}. A
theoretical understanding of this behavior is still lacking. 
In the present theoretical study we have studied the underlying mechanisms by
examining differences in the
alkali adsorption on ideal (1$\times$1) and (2$\times$1)MR reconstructed Cu(100) in
order to explain the specific behavior found for Li adsorption. 

\section{Computational Details}
Geometric structures of the Cu(100) substrate and of the
\mbox{Cu(100)+Ad}, \mbox{Ad = Li, Na, K} 
adsorbate systems are described in the repeated slab
geometry. The Cu substrate, in its unreconstructed (1$\times$1) and (2$\times$1)MR
reconstructed form (see Fig.~1), is approximated by 7 and 9 layer slabs
which are separated by vacuum corresponding to 5 substrate layers. This
has been found to yield a reliable surface representation \cite{Qua96}. The alkali
adsorbate layers (Li, Na, K) are added at the top and bottom of each
substrate slab where for coverages $\Theta = 0.25$ a lateral (2$\times$2) and
for  $\Theta = 0.5$ a c(2$\times$2) and (2$\times$1) 
supercell geometry is assumed as illustrated in
Fig.~2a--f. The geometries of the surface systems are further optimized by
allowing all atoms in the unit cells to relax according to the force field
obtained from the calculations described below. \\

The electronic structure of the surface systems is calculated within the
density functional theory (DFT) formalism using the local density
approximation (LDA) \cite{Cep80,Per81} as well as gradient corrected functionals
(GGA--II) \cite{Per92} for exchange and correlation. Total energies and
derived quantities are obtained with the full-potential linearized
augmented plane wave (FP-LAPW) method \cite{And75} implemented in 
the {\ttfamily WIEN93} program \cite{Bla93} where 45 k-points
inside the irreducible part of the Brillouin zone are used and the energy
cut--off for plane wave expansions in the interstitial region is set to 
$9.58$ Ry. Geometry optimizations of the surface systems are based on
calculations of forces acting on corresponding nuclei \cite{Koh96}
where equilibrium was assumed for forces below $3$ mRy/bohr.
A comparison of the geometric and energetic results from calculations using
the LDA and the GGA--II approach shows only small differences which do not
affect the present conclusions. Thus, we restrict ourselves in the
following to results from our LDA studies.

\section{Results and Discussion}
The (2$\times$1)MR reconstructed Cu(100) surface, shown in Fig.~1, is created
from the ideal unreconstructed surface by removing every second row of
surface atoms. As a result the top--most atoms of the reconstructed surface
experience a different electronic environment which affects the surface
geometry. This becomes obvious from Table 1a which gives numerical results
of the present geometry optimizations. For the unreconstructed Cu(100)
surface the interlayer spacing between the first and second surface layer,
\mbox{$d_{12}$ (= 1.76 Å} in the ideal bulk geometry), is decreased, 
$\Delta d_{12} = -3.2$\%,
and that between the second and third layer, $d_{23}$, slightly increased, 
$\Delta d_{23} = 0.2$\%, in good agreement with experimental \cite{Fow95} 
and other theoretical \cite{Rod93} results. 
In contrast, the (2$\times$1)MR reconstruction of pure Cu(100)
yields much larger changes in the interlayer spacings: $d_{12}$ is decreased by
almost $10$\% and $d_{23}$ by 2\%. 
In the present calculations the surface energy $\gamma$ is defined by
\begin{equation}
\gamma = [E_{\mathrm{tot}}(\mathrm{slab},n) - E_{\mathrm{tot}}(\mathrm{bulk},n)]/2
\end{equation}
where the total energy (per unit cell), $E_{\mathrm{tot}}(\mathrm{slab},n)$,
includes surface relaxation and reconstruction of the slab of $n$ layers 
while $E_{\mathrm{tot}}(\mathrm{bulk},n)$ is obtained for the
corresponding slab of the ideal bulk. For ideal (1$\times$1) Cu(100) the
calculations yield $\gamma = 0.8$ eV while for the (2$\times$1)MR reconstructed surface
one finds $\gamma = 1.0$ eV per (1$\times$1) unit cell. Thus, the MR reconstruction
requires $0.2$ eV per unit cell for Cu(100) which is about $10$ times larger
than the energy needed for the MR reconstruction of the Cu(110) surface
\cite{Vie96}. \\
 
Table 1a compares results from geometry optimizations on the different
\mbox{Cu(100)+Ad}, \mbox{Ad = Li, Na, K} adsorbate systems. Here the left part refers to
overlayer structures on unreconstructed Cu(100), (2$\times$2) for adsorbate
coverage $\Theta = 0.25$ (Fig.~2b) and c(2$\times$2) for $\Theta = 0.5$ (Fig.~2c) The right
part shows results for adsorption on (2$\times$1)MR reconstructed Cu(100), with
(2$\times$2) adsorbate geometry for $\Theta = 0.25$ (Fig.~2e) and with
(2$\times$1) for $\Theta = 0.5$ 
(Fig.~2f). Test calculations on Cu(100)+Li with different lateral
adsorbate sites showed that sites of maximum coordination close to Cu
substitutional sites were always energetically preferred in accordance
with Figs.~2b, c, e, f. Therefore, this lateral geometry was also assumed
for the Cu(100)+Na and Cu(100)+K systems. \\

For Cu(100)+Li the overlayer geometry (Figs.~2b, c) yields stable Li
centers at $z_{\mathrm{Cu-Li}} = 1.8$ Å above the substrate surface with only small
differences between coverages $\Theta_{\mathrm{Li}} = 0.25, 0.5$. This is in good agreement
with recent experimental LEED results, cp.~Table 1b. Further, substrate
relaxation, quantified by $\Delta d_{12}, \Delta d_{23}$, seems to be little influenced by
adsorption as evidenced from a comparison of the theoretical Cu(100)+Li
with the Cu(100) data. The overlayer
geometry results for Cu(100)+Na and Cu(100)+K are found to be rather
similar to those of Cu(100)+Li. Here the adsorbates stabilize for both
coverages at distances $z_{\mathrm{Cu-Na}} = 2.1$ Å and 
$z_{\mathrm{Cu-K}}=2.5$ Å above the
unreconstructed Cu(100) surface. The distance values are larger than
$z_{\mathrm{Cu-Li}}$ given above which is consistent with the increased atom size in
going from Li to Na to K. \\

Geometry differences between the different alkalis are much more
pronounced in the results for adsorption on the (2$\times$1)MR reconstructed
Cu(100) surface. For Cu(100)+Li the Li atoms arrange along the Cu troughs
(see Fig.~2e, f) stabilizing only slightly above the first Cu substrate
layer, \\ 
\mbox{$z_{\mathrm{Cu-Li}} = 0.4$ Å} for \mbox{$\Theta_{\mathrm{Li}} = 0.25$}, 
\mbox{$z_{\mathrm{Cu-Li}} = 0.2 - 0.3$ Å} for \mbox{$\Theta_{\mathrm{Li}} = 0.5$}.
The two different $z_{\mathrm{Cu-Li}}$
values given in Table 1a for $\Theta_{\mathrm{Li}} = 0.5$ refer to results of
adjacent Li centers in rows along the Cu troughs indicating a buckling of
the adsorbate layer. This
is reasonable due to the fact that the Cu--Cu nearest neighbor distance
which determines the lateral distance of the adsorbate rows is smaller
than typical Li metal nearest neighbor distances. The increased substrate
relaxation calculated for the pure (2$\times$1)MR reconstructed Cu(100) surface,
see above, exists also in the Cu(100)+Li adsorbate system with variations
of 3\%, see $\Delta d_{12}, \Delta d_{23}$ of Table 1a. A comparison of the theoretical
relaxation data with those of the LEED experiment on Cu(100)+Li \cite{Mey94}, 
cp.~Table 1b, yields good agreement. The geometries calculated for Na and K
adsorption on (2$\times$1)MR reconstructed Cu(100) differ considerably from those
of the Cu(100)+Li system. For the lower coverage, $\Theta = 0.25$,
the adsorbates stabilize at distances $z_{\mathrm{Cu-Na}} = 0.86$ Å and 
$z_{\mathrm{Cu-K}} = 1.7$ Å
above the substrate surface which is larger than the $z_{\mathrm{Cu-Li}}$ value and
explained by the different atom sizes. For $\Theta = 0.5$ the adsorbate layer
buckling is much more pronounced for the larger alkalis.  While in
Cu(100)+Li the hight difference between adjacent adsorbate centers is only
\mbox{$\Delta z_{\mathrm{Cu-Li}} = 0.11$ Å} the calculations 
yield \mbox{$\Delta z_{\mathrm{Cu-Na}} = 0.31$ Å} 
and \mbox{$\Delta z_{\mathrm{Cu-K}} = 2.31$ Å}. In fact, in Cu(100)+K the theoretical
buckling is so large that the
concept of a smooth adsorbate layer becomes questionable. \\

Fig.~3 shows color coded contour plots of the charge rearrangement in the
Cu(100)+alkali systems due to adsorption on the (2$\times$1)MR reconstructed
substrate. The rearrangement characterizes the electronic coupling of the
adsorbates with the substrate. It is quantified by the valence electron
difference between the adsorbate system and its separate components
\begin{equation}
\Delta \rho_{\mathrm val}({\bf r}) = \rho_{\mathrm val}(\mathrm{Cu(100)+Ad},{\bf r}) - 
\rho_{\mathrm val}(\mathrm{Cu(100)},{\bf r}) - \rho_{\mathrm val}(\mathrm{Ad},{\bf r})
\end{equation}
where positive $\Delta \rho_{\mathrm val}$ values indicate charge accumulation and negative
values depletion as a result of the adsorbate--substrate interaction. The
plots refer to Li (Fig.~3a), Na (Fig.~3b), and K (Fig.~3c) coverages 
$\Theta = 0.5$
and show $\Delta \rho_{\mathrm val}({\bf r})$ for a planar cut 
perpendicular to the surface along
the substrate troughs with Cu centers marked by circles and adsorbate
centers by squares. Obviously, the charge rearrangement leads, for all
three systems, to a depletion of charge above the (buckled) adsorbate
layer which is connected with the overall positive charging of the
adsorbates at the surface. The latter has been confirmed by charge
integration and is to be expected from simple electronegativity arguments.
Further, the plots show an almost homogeneous charge accumulation in the
region between adsorbate and substrate suggesting no directional
adsorbate--substrate bond formation. Thus, the alkali+Cu interaction is
determined to a major degree by electrostatic and metallic contributions.
Corresponding $\Delta \rho_{\mathrm val}({\bf r})$
plots for the adsorption on unreconstructed Cu(100)
lead to identical conclusions. These results confirm the complexity
of alkali adsorption on metal surfaces which has been discussed in 
the literature for a long time.  For reviews see Refs.~\cite{Bon89,Sta95}. \\ \\
The relative stability of the Cu(100)--alkali systems for adsorption on the
(2$\times$1)MR reconstructed vs. the unreconstructed Cu(100) substrate can be
determined from total energy studies. For this purpose we define a
reconstruction energy $E_s(\Theta)$ by
\begin{equation}
E_s(\Theta) = \gamma_{\mathrm{rec}} - 
\gamma_{\mathrm{unrec}} + 
D_{\mathrm{rec}}(\Theta) -
D_{\mathrm{unrec}}(\Theta) 
\end{equation}
where $\gamma_{\mathrm{rec}},\gamma_{\mathrm{unrec}} $ 
denote surface energies of the pure substrate with and
without reconstruction while 
$D_{\mathrm{rec}}(\Theta), D_{\mathrm{unrec}}(\Theta) $ are adsorbate binding
energies for the two substrate geometries obtained from respective total
energies. Thus, $E_s(\Theta)$ includes the energy required to reconstruct the
substrate as well as the difference of adsorbate-binding with and without
reconstruction. Positive $E_s(\Theta)$ values suggest that the
unreconstructed adsorbate system is energetically preferred while negative
values yield the reconstructed system to be favored. Fig.~4 shows the
result of the present calculations in a level diagram of $E_s(\Theta)$ for
adsorbate coverages \mbox{$\Theta = 0, 0.25, 0.5$} ($\Theta= 0$ denotes the pure substrate). On
pure Cu(100) the reconstruction energy amounts to $E_s = 0.2$ eV per
unit cell. This energy is reduced to $0.03$ eV for Li and Na and to $0.05$ eV
for K adsorption at coverages \mbox{$\Theta = 0.25$} which demonstrates that the
adsorbates are always bound more strongly to the (2$\times$1)MR reconstructed
than to the unreconstructed Cu(100) substrate. At an increased coverage 
\mbox{$\Theta = 0.5$} the reconstruction energy is further decreased to $0.02$ eV for Li
adsorption. In fact, a conservative estimate of numerical errors in the
calculations, yielding $\pm 0.02$ eV, suggests that for Cu(100)+Li the
reconstructed and unreconstructed systems become equally stable 
(\mbox{$E_s \approx 0$ eV})
for $\Theta = 0.5$. In contrast, the reconstruction energy leads to positive
values for the two larger alkali adsorbates, \mbox{$E_s(\Theta=0.5) = 0.14$ eV} 
for Na and \mbox{$E_s(\Theta=0.5)= 0.19$ eV} for K. This indicates that for
Cu(100)+Na and Cu(100)+K
adsorption on the unreconstructed substrate is energetically preferred
which confirms the experimental findings \cite{Oka91,Aug85,Mey94}.

\section{Conclusions}
The present DFT total energy studies reveal interesting details of the
adsorbate induced substrate reconstruction in the 
\mbox{Cu(100)+alkali}
systems. Geometry optimizations show that in Cu(100)+Li the adsorbate
can stabilize on the (2$\times$1) MR reconstructed surface above the substrate
troughs where the adsorbate-substrate binding is stronger compared to
adsorption on the unreconstructed surface. As a result, the adsorbate
systems with and without substrate reconstruction become equally stable
for coverages $\Theta = 0.25 - 0.5$ and transitions between the two becomes
possible with varying temperature confirming the experimental findings
\cite{Toc92,Miz93}. In contrast, calculations on the Cu(100)+Na and
Cu(100)+K systems show that, for coverages $\Theta = 0.25 - 0.5$, adsorption on
the unreconstructed Cu(100) surface is always energetically preferred over
that on a (fictitious) (2$\times$1) MR reconstructed surface, again confirming
the experiment \cite{Oka91,Aug85,Mey94}. Based on the calculations, this different
behavior between the alkali adsorbates can be explained in parts by simple
geometric effects. The larger alkalis, Na, K, cannot stabilize close
enough to the reconstructed substrate surface and their atom sizes lead to
strong layer buckling for higher adsorbate coverages which makes
adsorption on the reconstructed substrate very unlikely to occur.
It should be emphasized that the present calculations refer to static 
equilibrium structures of Cu(100)+alkali which can be compared for different
constraints, overlayer or missing row. Transition states between the different
structures and respective energy barriers require the knowledge of appropriate
reaction paths which have to be assumed or obtained from separate ab initio
molecular dynamics calculations. 
\newpage
\parindent0mm
{\Large \bf Acknowledgment}
\\ \\ 
This study was supported in parts by Deutsche Forschungsgemeinschaft
(Sonderforschungsbereich 290, Berlin).
\newpage
\bibliographystyle{prsty}
\bibliography{paper}
\parindent0mm
\newpage
\ \\
{\Large \bf Tables}
\\ \\
Table 1. \\
Geometry results of the Cu(100) substrate and of the
\mbox{Cu(100)+Ad}, \\ 
\mbox{Ad = Li, Na, K} adsorbate systems with and without substrate
reconstruction and for adsorbate coverages $\Theta = 0.25 - 0.5$. Table 1a lists
theoretical data from the present LDA calculations and from those of 
ref.~\cite{Rod93}
while Table 1b compares experimental data of refs.~\cite{Miz93},
\cite{Mey94}, \cite{Rod93}, \cite{Fow95}, \cite{Miz93b}. For definitions
of the (differential) perpendicular distances, 
$z_{\mathrm{Cu-Ad}}, \Delta d_{12}, \Delta d_{23}$, see text. 
\newpage
\begin{sidewaystable}
\centering
\begin{tabular}{|l@{}@{}l|ccc|ccc|} \hline
& & \multicolumn{2}{l}{\bf Overlayer structure} & & 
\multicolumn{3}{l|}{\bf Missing row reconstruction} \\ \hline
& Coverage $\Theta$ (ML)& z$_{\mbox{\footnotesize Cu-Ad}}$ (\AA) & 
$\Delta$d$_{12}$ (\%) & $\Delta$d$_{23}$ (\%) & 
z$_{\mbox{\footnotesize Cu-Ad}}$ (\AA) & 
$\Delta$d$_{12}$ (\%) & $\Delta$d$_{23}$ (\%) \\ \hline \hline
{\bf (a) Theory:} & & & & & & &  \\ \hline
{\bf Cu(100)} & & & & & & &   \\
& $0.0$ & -- &  $-3.2$  & $+0.2$ & -- & $-9.6$  & $-2.1$ \\
& $0.0$ \cite{Rod93} & -- & $-3.0$ & $+0.1$ & -- & --  & -- \\
{\bf Cu(100)+Li}  & & & & & & &  \\
& $0.25$ & $1.82$ & $-4.0$  & $-2.7$ &
$0.42$ & $-12.2$ & $-3.4$  \\ 
& $0.50$ & $1.81$ & $-2.7$  & $-1.9$ & 
$0.20/0.31$ & $-8.2$  & $-1.8$  \\ 
{\bf Cu(100)+Na} &  & & & & & &  \\
& $0.25$ & $2.09$  &  $-4.8$  & $-4.3$ &
$0.86$ & $-13.1$  & $-3.8$ \\
& $0.50$ & $2.12$ & $-4.1$ & $-4.0$  &
$0.62/0.93$ & $-9.2$  & $-3.5$  \\
{\bf Cu(100)+K} & & & & &  & &  \\
& $0.25$ & $2.48$ & $-4.8$ & $-3.2$ &
$1.27$ & $-9.3$ & $-3.5$ \\
& $0.50$ & $2.56$ & $-4.3$ & $-4.7$ &
$1.16/3.47$ & $-9.9$  & $-3.4$  \\ \hline
{\bf (b) Experiment:} & & & & & & &  \\ \hline
{\bf Cu(100)} & & & & & & &  \\
& $0.0$ \cite{Fow95} & -- &  $-2.0 \pm 0.2$ & $+0.5 \pm 1.2$ & -- & --  & -- \\
{\bf Cu(100)+Li} & & & & & & &  \\
& $0.25-0.55$ \cite{Miz93,Miz93b} & $1.96 \pm 0.08$ & $0 \pm 2$ & -- & 
 -- & $-7 \pm 3$ & $0 \pm 2$ \\
{\bf Cu(100)+K} & & & & & & &  \\
& $0.25$ \cite{Mey94} & $2.25 \pm 0.15$ & -- & -- & -- & -- & -- \\ \hline
\end{tabular}
\end{sidewaystable}
\clearpage
\newpage
{\Large \bf Figure Captions}
\\ \\
Fig.~1 \\ 
Perspective view of the Cu(100) surface showing the ideal (1$\times$1)
and the (2$\times$1) missing row reconstructed geometry. The two--dimensional
elementary cells are marked by rectangles for each geometry.
\\ \\
Fig.~2 \\  Top view of the pure Cu(100) and of the adsorbate covered
\mbox{Cu(100)+alkali} surfaces illustrating different surface
geometries considered in the present study. The Cu substrate atoms are
shown as dark (1st layer) and light (2nd, 3rd layer) gray balls while the
alkali adsorbate atoms are indicated by red balls. The upper part of 
Fig.~2 refers to adsorption on the unreconstructed Cu(100) surface showing (a)
the ideal (1$\times$1) substrate, (b) (2$\times$2) adsorption 
($\Theta = 0.25$), and  (c)
c(2$\times$2) adsorption ($\Theta = 0.5$). The lower part refers to adsorption on the
reconstructed surface showing (d) the (2$\times$1) missing row substrate, (e)
(2$\times$2) adsorption ($\Theta = 0.25$), and 
(c) (2$\times$1) adsorption ($\Theta = 0.5$).
\\ \\
Fig.~3 \\
Color coded contour plots of the charge density difference,
$\Delta \rho_{\mathrm val}({\bf r})$, of Cu(100) covered with alkali adsorbates (a) Li, (b) Na, and
(c) K. The plots refer to the (2$\times$1) missing row reconstructed substrate
and adsorbate coverages $\Theta = 0.5$ and show $\Delta \rho_{\mathrm val}({\bf r})$ for a planar cut
perpendicular to the surface along the substrate troughs. For a definition
of $\Delta \rho_{\mathrm val}({\bf r})$ see text. The units used for the color coding, see right
scale, are 1/bohr$^3$.
\\ \\
Fig.~4 \\
Level diagram of the reconstruction energy $E_s(\Theta)$ for the
\mbox{Cu(100)+Ad}, \mbox{Ad = Li, Na, K} systems at adsorbate coverages $\Theta = 0, 0.25, 0.5$
($\Theta = 0$ denotes the pure substrate). For a definition of $E_s$ see text. The
error bar of $\pm 0.02$ eV shown with the Li result for $\Theta = 0.5$ gives an
estimate of numerical errors in the calculations.
\newpage
\pagestyle{empty}
\begin{figure}[ht]
\begin{center}
\begin{turn}{270}
\epsfig{file=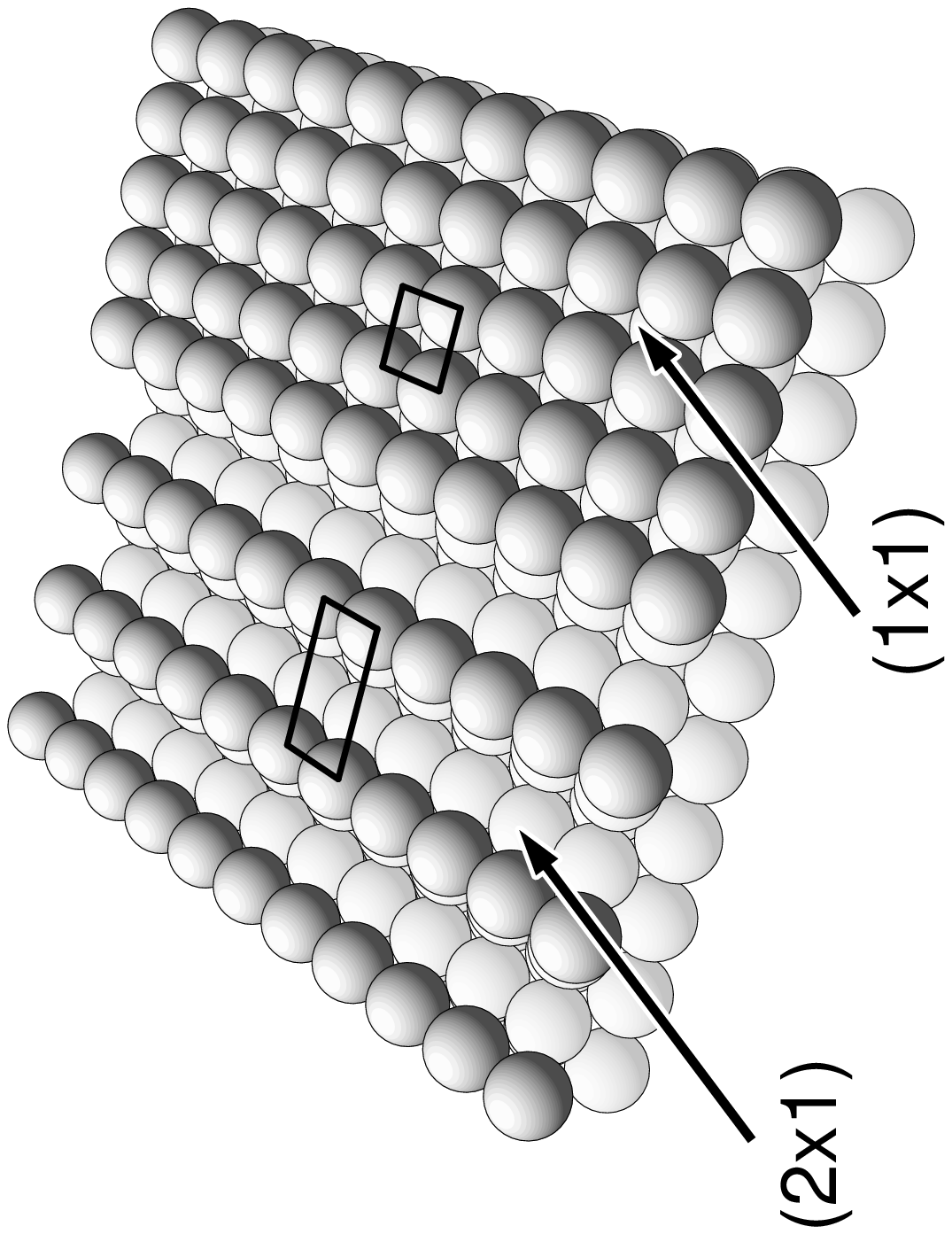,scale=1.0}
\end{turn}
\end{center}
\label{fig:subst}
\end{figure}
\newpage
\setlength{\hoffset}{-10mm}
\begin{figure}[ht]
\begin{center}
\begin{turn}{270}
\epsfig{file=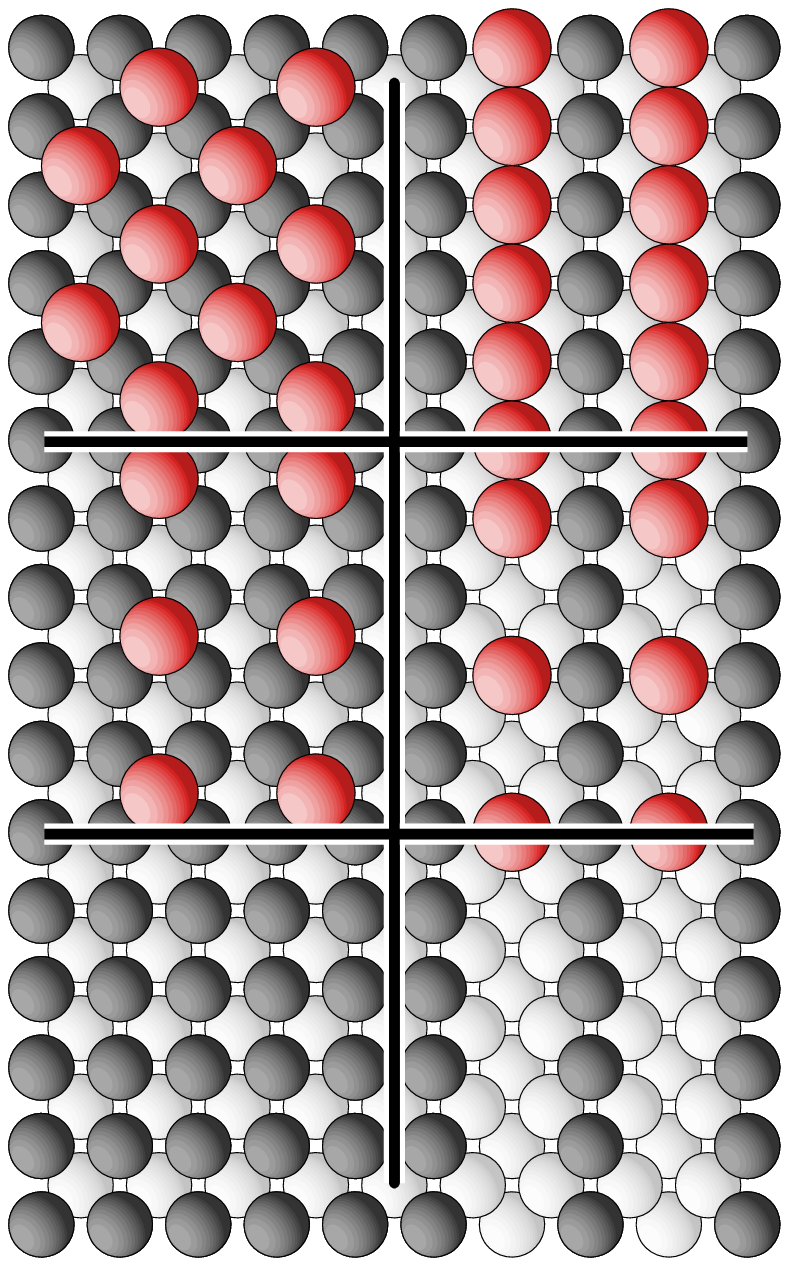,scale=1.3}
\end{turn}
\end{center}
\label{fig:motiv}
\end{figure}
\pspicture(0,0)(21,3)
\rput{0}(1,15.5){\huge \sf (a)}
\rput{0}(6,15.5){\huge \sf (b)}
\rput{0}(11,15.5){\huge \sf (c)}
\rput{0}(1,2.5){\huge \sf (d)}
\rput{0}(6,2.5){\huge \sf (e)}
\rput{0}(11,2.5){\huge \sf (f)}
\endpspicture
\newpage
\setlength{\hoffset}{100mm}
\begin{figure}[ht]
\begin{turn}{90}
\begin{minipage}[b]{0.43\linewidth}
\begin{turn}{270}
\centering\epsfig{file=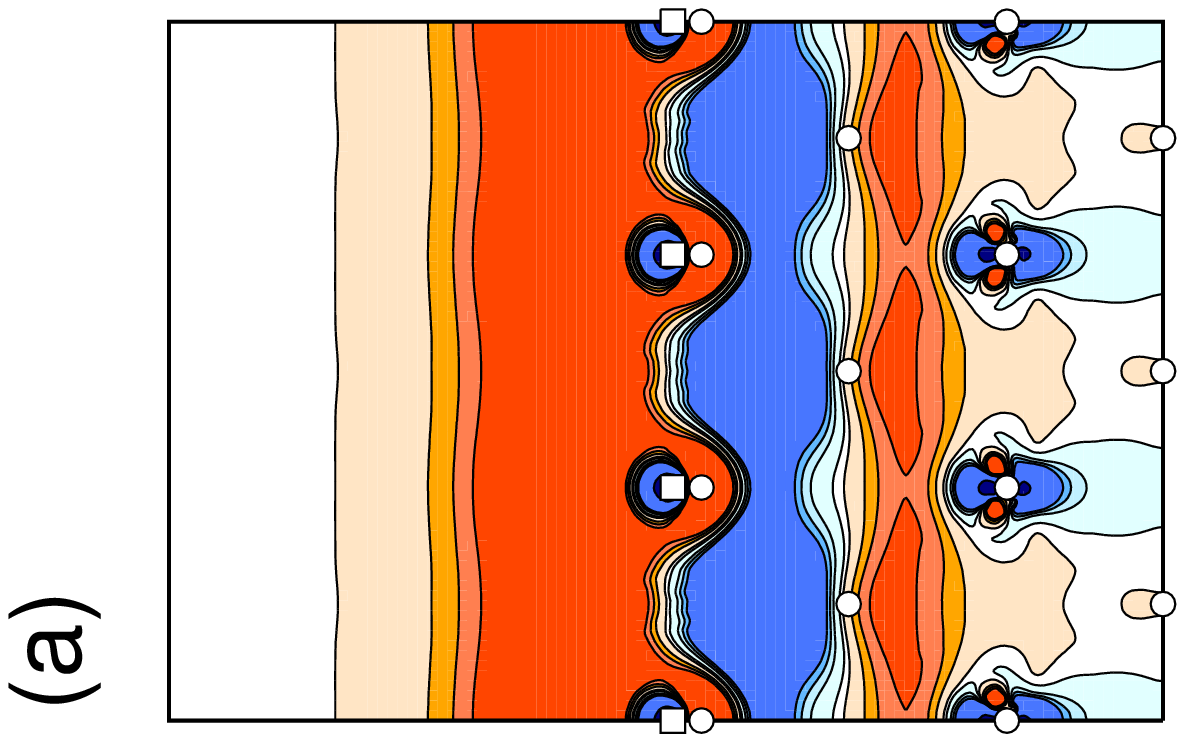,scale=0.8}
\end{turn}
\end{minipage}
\begin{minipage}[b]{0.43\linewidth}
\begin{turn}{270}
\centering\epsfig{file=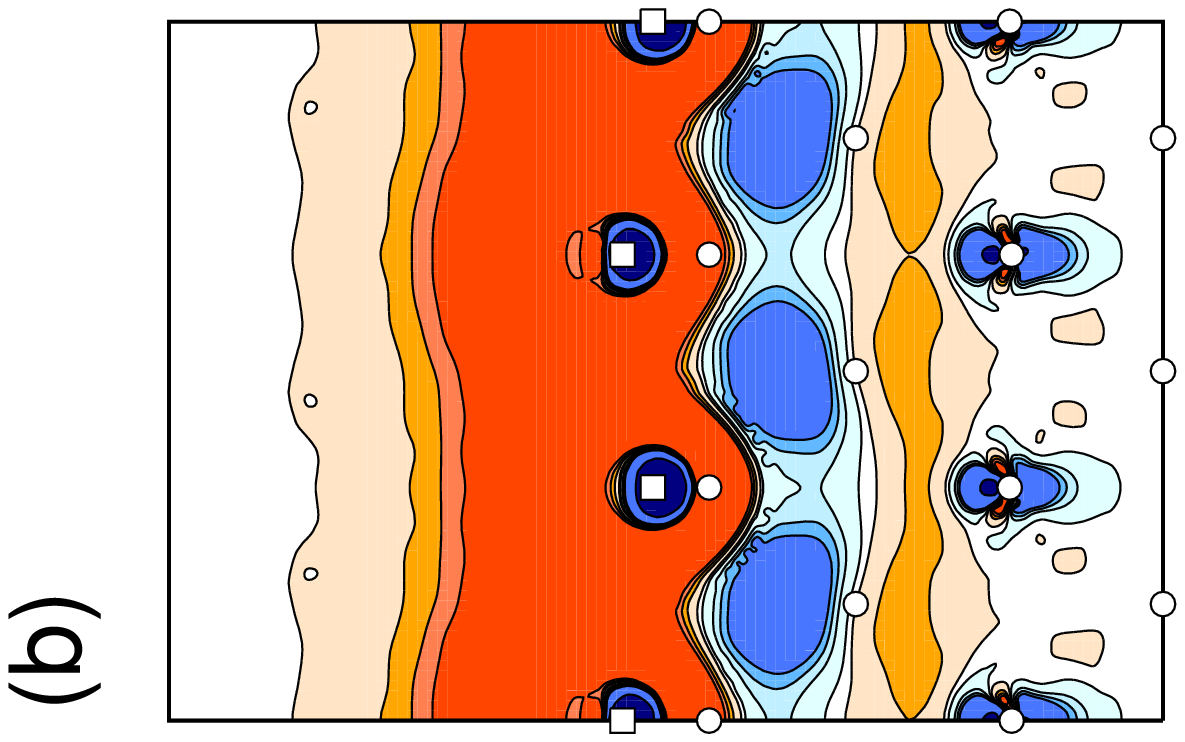,scale=0.8}
\end{turn}
\end{minipage}
\begin{minipage}[b]{0.43\linewidth}
\begin{turn}{270}
\centering\epsfig{file=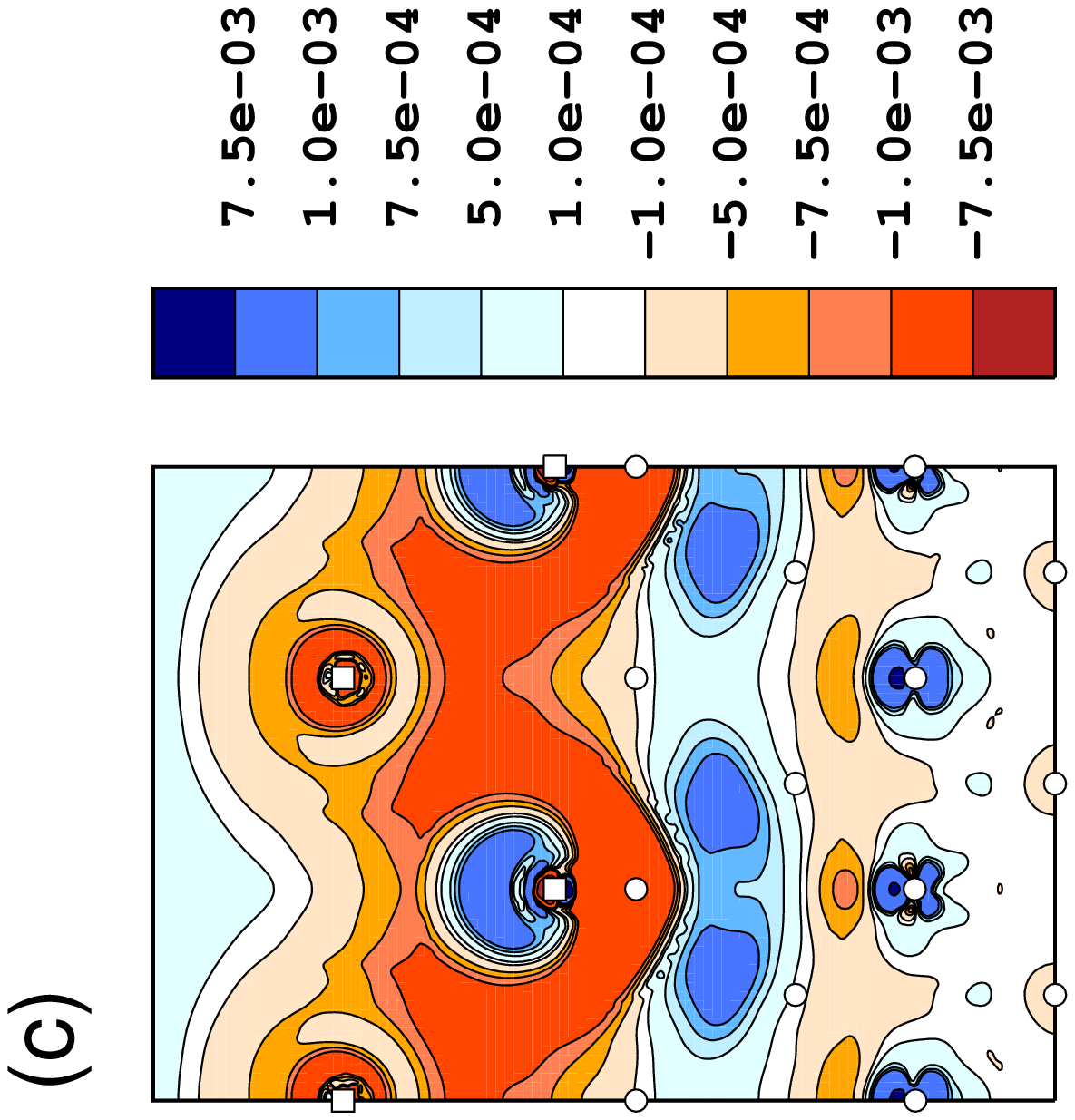,scale=0.8}
\end{turn}
\end{minipage}\hfill
\end{turn}
\label{fig:charge}
\end{figure}
\clearpage
\newpage
\setlength{\hoffset}{1cm}
\psset{unit=1.0cm} 
\pspicture(0,0)(15,13)
\newgray{lgray}{0.90}
\psframe[linewidth=0.0,fillstyle=solid,fillcolor=lgray](0,-2)(13,0)
\psline[linewidth=0.07,linestyle=dashed]{c-c}(0,0)(13,0)
\psline[linewidth=0.12]{c-c}(0,-2)(13,-2)
\psline[linewidth=0.12]{c-c}(0,-2)(0,12)
\psline[linewidth=0.12]{c-c}(13,12)(13,-2)
\psline[linewidth=0.12]{c-c}(13,12)(0,12)
\rput{0}(7,-4.5){\huge \bf \sf $\Theta$ (monolayer)}
\rput{90}(-3,5.5){\huge \bf \sf E$_{\sf{s}}$ (eV)}
%
\psline[linewidth=0.1](0,-1)(0.25,-1)
\psline[linewidth=0.1](0,0)(0.5,0)
\psline[linewidth=0.1](0,1)(0.25,1)
\psline[linewidth=0.1](0,2)(0.25,2)
\psline[linewidth=0.1](0,3)(0.25,3)
\psline[linewidth=0.1](0,4)(0.25,4)
\psline[linewidth=0.1](0,5)(0.5,5)
\psline[linewidth=0.1](0,6)(0.25,6)
\psline[linewidth=0.1](0,7)(0.25,7)
\psline[linewidth=0.1](0,8)(0.25,8)
\psline[linewidth=0.1](0,9)(0.25,9)
\psline[linewidth=0.1](0,10)(0.5,10)
\psline[linewidth=0.1](13.0,-1)(12.75,-1)
\psline[linewidth=0.1](13.0,0)(12.75,0)
\psline[linewidth=0.1](13.0,1)(12.75,1)
\psline[linewidth=0.1](13.0,2)(12.75,2)
\psline[linewidth=0.1](13.0,3)(12.75,3)
\psline[linewidth=0.1](13.0,4)(12.75,4)
\psline[linewidth=0.1](13.0,5)(12.5,5)
\psline[linewidth=0.1](13.0,6)(12.75,6)
\psline[linewidth=0.1](13.0,7)(12.75,7)
\psline[linewidth=0.1](13.0,8)(12.75,8)
\psline[linewidth=0.1](13.0,9)(12.75,9)
\psline[linewidth=0.1](13.0,10)(12.5,10)
\rput{0}(-1.5,0){\huge \sf 0.00}
\rput{0}(-1.5,5){\huge \sf 0.10}
\rput{0}(-1.5,10){\huge \sf 0.20}
%
\psline[linewidth=0.1](5,-2.0)(5,-1.5)
\psline[linewidth=0.1](10,-2.0)(10,-1.5)
\rput{0}(0,-3.0){\huge \sf 0.00}
\rput{0}(5,-3.0){\huge \sf 0.25}
\rput{0}(10,-3.0){\huge \sf 0.50}
\psline[linewidth=0.1](0,10)(2,10)
\rput{0}(4,10){\huge \sf Cu(100)} 
\psline[linewidth=0.1](4,2.3)(6,2.3)
\rput{0}(2.5,2.5){\huge \sf K}
\psline[linewidth=0.1](4,1.6)(6,1.6)
\rput{0}(2.5,1.5){\huge \sf Li,Na}
\psline[linewidth=0.1](4,1.6)(6,1.6)
\psline[linewidth=0.04,linestyle=dashed](2,10)(4,2.3)
\psline[linewidth=0.04,linestyle=dashed](2,10)(4,1.6)
\psline[linewidth=0.1](9,9.5)(11,9.5)
\rput{0}(12,9.5){\huge \sf  K}
\psline[linewidth=0.1](9,7.3)(11,7.3)
\rput{0}(12,7.3){\huge \sf  Na}
\psline[linewidth=0.1](9,1.05)(11,1.05)
\rput{0}(12,1.05){\huge \sf  Li}
\psline[linewidth=0.05](10,2.3)(10,-0.2)
\psline[linewidth=0.05](9.75,2.3)(10.25,2.3)
\psline[linewidth=0.05](9.75,-0.2)(10.25,-0.2)
\psline[linewidth=0.04,linestyle=dashed](6,2.3)(9,9.5)
\psline[linewidth=0.04,linestyle=dashed](6,1.6)(9,7.3)
\psline[linewidth=0.04,linestyle=dashed](6,1.6)(9,1.05)
\endpspicture
\ \vspace*{4cm}
\end{document}